\documentclass[twocolumn,twoside,slac_two]{revtex4}
\usepackage{graphicx}
\usepackage{fancyhdr}
\begin{document}

\title{Multiwavelength campaign of the gamma-ray flaring source PKS\,2052$-$47}


\author{C. S. Chang$^{1}$ }

\author{E. Ros$^{1,2}$}

\author{M. Kadler$^{3,4,5}$}

\author{R. Ojha$^{6,7}$}

\author{the \textit{Fermi} LAT Collaboration}

\author{the TANAMI team}

\author{the F-GAMMA team\footnote{The F-GAMMA team (in alphabetical order): E. Angelakis, L. Fuhrmann, T. P. Krichbaum, S. Larsson, N. Marchili, G. Nestoras, H. Ungerechts, and J. A. Zensus et al.} }

\affiliation{$^{1}$Max-Planck-Institut f\"ur Radioastronomie, Auf dem H\"ugel 69, D-53121 Bonn, Germany}
\affiliation{$^{2}$Departament d'Astronomia i Astrof\'{\i}sica, Universitat de Val\`encia, E-46100 Burjassot, Spain }
\affiliation{$^{3}$Dr. Remeis-Sternwarte \& ECAP, Sternwartstr. 7, D-96049 Bamberg, Germany }
\affiliation{$^{4}$CRESST/NASA Goddard Space Flight Center, Greenbelt, MD 20771, USA}
\affiliation{$^{5}$USRA, 10211 Wincopin Circle, Suite 500 Columbia, MD 21044, USA}
\affiliation{$^{6}$USNO, 3450 Massachusetts Ave., NW, Washington DC 20392, USA }
\affiliation{$^{7}$NVI, Inc., 7257D Hanover Parkway, Greenbelt, MD 20770, USA }

\begin{abstract}
The flat-spectrum radio quasar PKS\,2052$-$47 experienced an optical flare in July$-$August, 2009. On August 9, the LAT detector onboard \textit{Fermi} observed a flare from the source, which reached its active phase in the following two months. To further investigate the physics and emission mechanisms during the flaring state of this source, we arranged a multiwavelength campaign from radio to gamma-ray to study its spectral energy distribution (SED) and its brightness variations across the spectrum. Here we present the first results of this multiwavelength campaign.

\end{abstract}

\maketitle

\thispagestyle{fancy}

\section{Introduction}
\begin{figure*}[Ht]
 \centering
 %% 1st image
 \begin{minipage}{.48\textwidth}%[!b]
   \centering
   \includegraphics[width=0.98\textwidth]{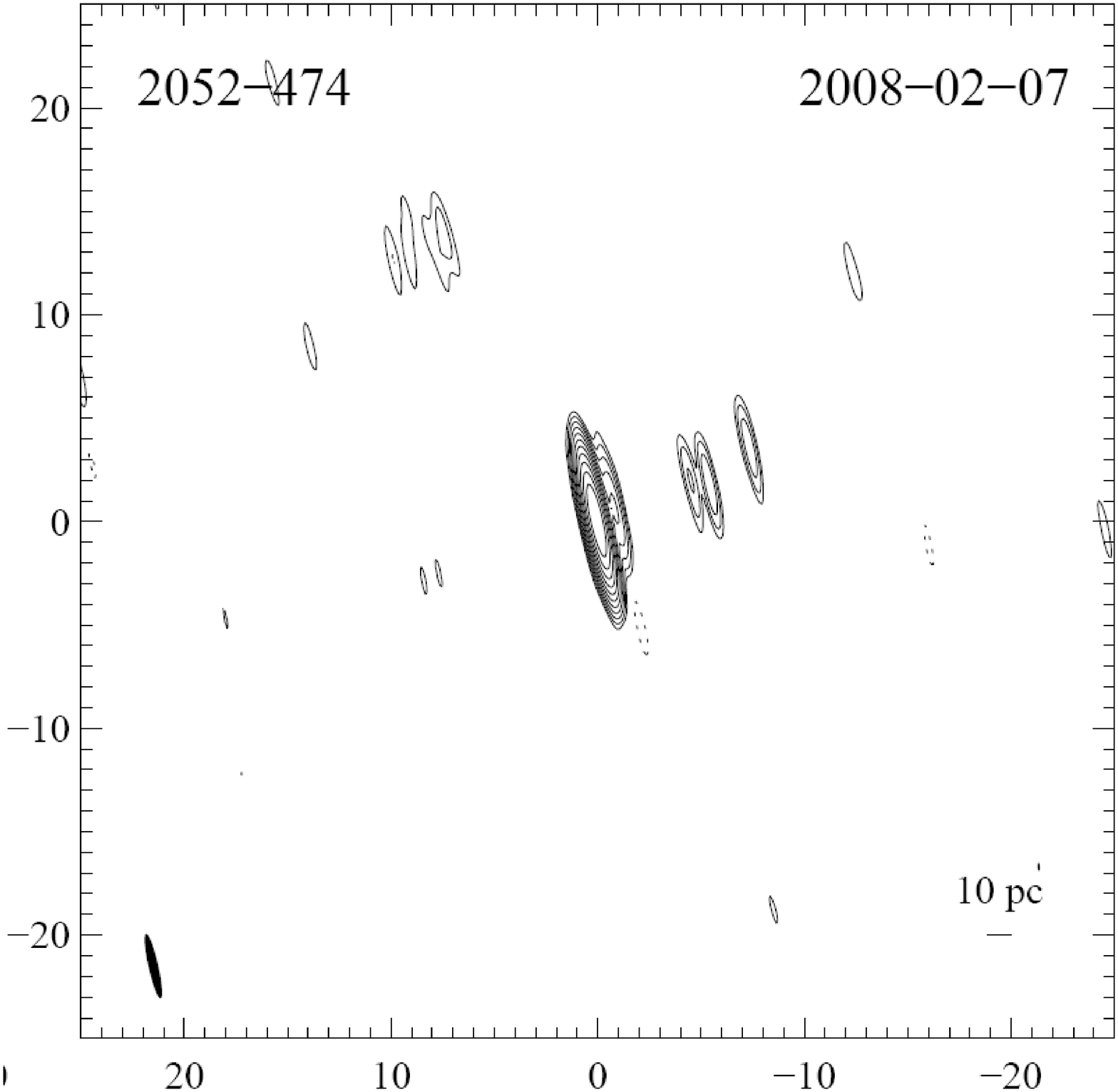}
 \end{minipage}
 \hspace{0.05cm}
 %% 2nd image
 \begin{minipage}{.48\textwidth}%[!b]
  \centering
  \includegraphics[width=0.98\textwidth]{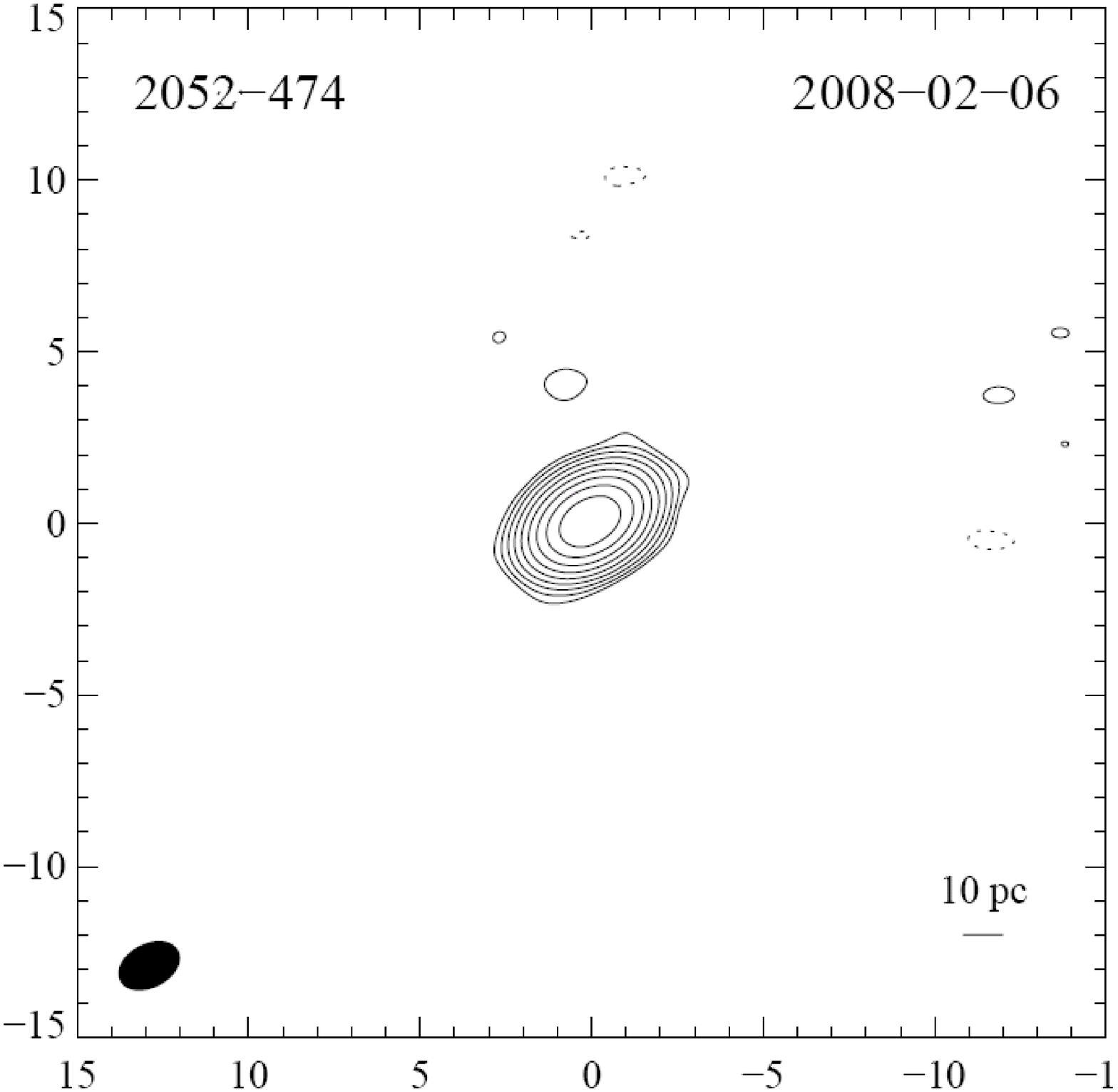}
 \end{minipage}
 %% caption
 \begin{minipage}{0.98\textwidth}%[t]
 \caption{The TANAMI VLBI images of PKS\,2052$-$47 at 8\,GHz (left, February 7, 2008) and 22\,GHz (right, February 6, 2008). The axes are relative right ascension (RA) and declination (DEC) in milli-arcsecond (mas). On the bottom right, the scale of 10 parsec (pc) is shown. The scales are different in both images. }
 \label{fig:tanami}
 \end{minipage}

\end{figure*}

\vspace{1mm}
%\smallskip
%\setlength{\parskip}{0cm}
PKS\,2052$-$47 is a flat-spectrum radio quasar with a redshift of 1.489 \cite{jau84}. Earlier Australia Telescope Compact Array (ATCA) and \textit{Chandra} X-ray observations showed that the source has a two-sided jet at kiloparsec scales in the radio band, with no extended emission in X-rays \cite{mar05}. Very Large Baseline Interferometry (VLBI) observations by TANAMI\footnote{\texttt{http://pulsar.sternwarte.uni-erlangen.de/tanami/}} (Tracking Active Galactic Nuclei with Austral Milliarcsecond Interferometry, see \cite{ojh08,kad07,boe09}) show that the parsec-scale structure is very compact, with a high brightness-temperature core of $2 \times 10^{12}$\,K and a size of $(0.3\times 0.1)$\,mas at 8.4\,GHz. At 22\,GHz the source is unresolved with a restoring beam of $(1.9 \times 1.3)$\,mas. At 8.4\,GHz, a very faint jet to the west is seen (see Figure \ref{fig:tanami}).

\medskip

From July 2009, the Automatic Telescope for Optical Monitoring (ATOM) team reported a steady increase of flux density from PKS\,2052-47, which reached its high state and flared in the optical band in August \cite{hau09}. Right after this, the \textit{Fermi} Large Area Telescope (LAT) observed a gamma-ray flare from this source \cite{cha09}. To understand the physics at work and emission mechanisms during its flaring state, we arranged a multi-wavelength campaign from radio to gamma-ray in early September 2009 in order to follow its active phase.    %Following its flaring state, we will investigate the correlations of variability between differen wavebands, and to understand the nature of this source.

\bigskip
\bigskip

\section{Multiwavelength observations}
The multiwavelength campaign of PKS\,2052$-47$ was arranged simultaneously with the scheduled southern VLBI observations of TANAMI during early September 2009. The observation log is reported in Table \ref{tab:obs_log}. The participating facilities include: Ceduna-Hobart Interferometer (CHI), the large APEX bolometer camera (LABOCA) \cite{sir09} at the Atacama Pathfinder EXperiment (APEX, in the framework of the F-GAMMA project, see \cite{fuh07}),  the Ultra-Violet/Optical Telescope (UVOT) and X-Ray Telescope (XRT) onboard \textit{Swift}, and the \textit{Fermi} Large Area Telescope. The \textit{Swift} target-of-opportunity observations covered five days centered at the date of the TANAMI observations, and the exposure time for each observation was between 1500 and 2500 seconds.

%%%%%%%%%%%%%%%%%(22)
%%%%%%%%%%%%%%%%%%%%%

%% Observation log ---------------------------------------
\begin{table*}%[Hb]%[!t]
%\centering
\caption{Observation log}
\begin{tabular*}{0.778\textwidth}{@{}lllll@{}}
%\hline
\hline
%\multicolumn{2}{c}{Epoch}        &  \multicolumn{2}{c}{S$_{\mathrm{\nu}}$ \footnotesize{[mJy]}} & $\alpha$  \\
%$\lambda$2\,cm & $\lambda$20\,cm &  $\lambda$2\,cm & $\lambda$20\,cm &           \\
\textbf{Start date} & \textbf{End date} & \textbf{Facility} &  \textbf{Band} & \textbf{Comments} \\
\hline
2009-09-05 & 2009-09-06 &  CHI & 6.6 GHz   & -  \\
%\hline
2009-09-05 & 2009-09-06 &  TANAMI        & 8 \& 22 GHz       & -  \\
%\hline
2009-09-02 & 2009-09-06 &  APEX          & 345 GHz           & -  \\
%\hline
2009-09-04 & 2009-09-08 &  \textit{Swift} UVOT    & 2000--3500 \AA      & $\sim$2\,ks per day   \\
%\hline
2009-09-04 & 2009-09-08 &  \textit{Swift} XRT     & 0.3--10 keV      & $\sim$2\,ks per day  \\
%\hline
2008-06-11 & present    &  Fermi LAT     & 100 MeV--300 GeV & All-sky survey since June 11, 2008 \\
\hline
%& & & & \\
\end{tabular*}
\label{tab:obs_log}
\end{table*}
%%------------------------------------------------

\renewcommand{\dbltopfraction}{0.9}

%%%%%%%%%%%%%%%%(29)
\begin{figure*}[p]
 \centering
 %% 1st image
 \begin{minipage}{0.46\textwidth}%[!b]
   \centering
   \includegraphics[angle=270, width=1\textwidth]{fig_xrt_xspec.ps}
   \caption{\textit{Swift} XRT spectrum of PKS\,2052$-$47 in the 0.3$-$6 keV energy range with 2.5 ks integration observed on September 6, 2009. The fitted model (solid line, upper panel) is an absorbed power-law. The residual of fit is shown in the lower panel.\\ \bigskip }
   \label{fig:xrt_spec}
 \end{minipage}
 \hspace{1cm}
 %% 2nd image
 \begin{minipage}{0.46\textwidth}%[!b]
  \centering
  \includegraphics[width=1\textwidth]{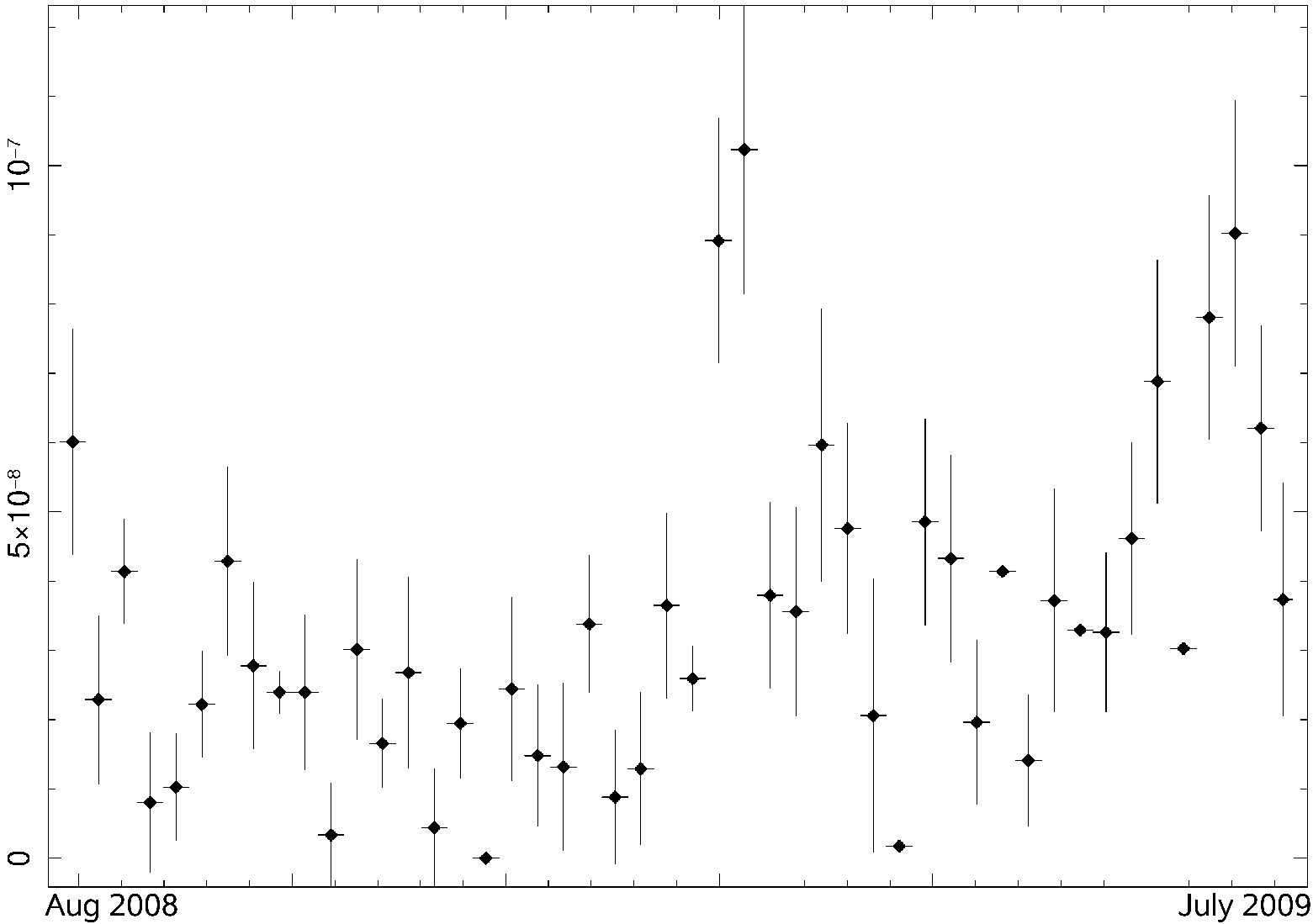}
  \caption{\textit{Fermi} LAT pre-flare light curve of PKS\,2052$-$47 from 4$^{\mathrm{th}}$ August 2008 to 4$^{\mathrm{th}}$ July 2009. This plot shows the light curve since Fermi's launch, binned to one week. The x-axis is time in the unit of mission elapsed time (MET), and the y-axis is the energy flux between 100\,MeV--100\,GeV in units of photon cm$^{-2}$ s$^{-1}$. }
  \label{fig:lat_lc}
 \end{minipage}
 %%3rd
 \begin{minipage}{0.98\textwidth}%[!b]
  \centering
  \includegraphics[angle=270, width=1\textwidth]{fig_sed.ps}
  \caption{The broadband SED of PKS\,2052$-$47. Black filled pentagons: simultaneous data from our multiwavelength campaign triggered in September 2009, which includes APEX preliminary data, \textit{Swift} UVOT/XRT, and \textit{Fermi} LAT data. Red open circles: historical data from the NASA extragalactic database (NED), including the EGRET detection. The highest energy bin of LAT data (right-most filled pentagon) has an error bar which has the same order of magnitude as the flux density value.} 
 \label{fig:sed}
 \end{minipage}

\end{figure*}
%%%%%%%%%%%%%%%%%%%%%%%5

%%%%%%%%%%%%%%%%%%%%%%(7)
%\begin{figure*}[Hb]
% \includegraphics[angle=270, width=1\textwidth]{fig_sed.ps}
% \caption{The broadband SED of PKS\,2052$-$47. Black filled pentagons: simultaneous data from the multiwavelength campaign triggered in September 2009, which includes APEX preliminary data, \textit{Swift} UVOT/XRT, and \textit{Fermi} LAT data. Red open circles: historical data from the NASA extragalactic database (NED), including the EGRET detection.  }
% \label{fig:sed}
%\end{figure*}
%%%%%%%%%%%%%%%%%%%%%%

\section{Preliminary results}

As of December 2009, the data collected include: \textit{Swift}, \textit{Fermi} LAT, and APEX preliminary results (see Table \ref{tab:obs_log}). The preliminary APEX flux density presented here was provided by the F-GAMMA team, and the final data calibration is in process. The TANAMI and CHI simultaneous datasets will be correlated shortly. To illustrate the radio morphology at parsec-scales, we show two TANAMI VLBI images of PKS 2052$-$47 obtained in February 2008 at 8 and 22\,GHz (Ojha et al., submitted to A\&A, and Kadler et al. in preparation) in Figure \ref{fig:tanami}. Notice that the image resolution of the 8\,GHz map is higher than that at 22\,GHz, due to the inclusion of a telescope in Chile, equipped only with a receiver at 8\,GHz, providing longer baselines to better resolve the source. 

We show a \textit{Swift}/XRT spectrum in the 0.3$-$6 keV energy range with 2.5 ks integration in Figure \ref{fig:xrt_spec}, taken on September 6, 2009. We used \texttt{XSPEC} version 12.5 to fit an absorbed power-law to the spectrum with a fixed Galactic neutral hydrogen column density of 2.79$\times10^{20}$ cm$^{-2}$ \cite{kal05}, and obtained a photon index of 1.53 $\pm$ 0.11. The value of the Cash statistic after fitting is 21.2 with 20 degrees of freedom. The photon statistics above 6 keV were not sufficient to obtain spectral data. 

We show a weekly-binned \textit{Fermi} LAT light curve (100\,MeV$-$100\,GeV) from August 2008 to July 2009 in Figure \ref{fig:lat_lc}, which is the first 11-month of the mission. We used \textit{Fermi} Science Tools version v9r10 and instrumental response functions (IRFs) version \texttt{P6\_V3\_DIFFUSE} to analyze LAT data. From Figure \ref{fig:lat_lc}, we could see that there is a gradual rise in the flux density with time, especially when time approached to July 2009. 

We present a preliminary broadband SED of PKS\,2052$-$47 from radio to gamma-ray in Figure \ref{fig:sed}. As shown in Figure \ref{fig:sed}, the synchrotron self-Compton (SSC) peak appears to be higher than the synchrotron emission peak in our result. This is consistent with the results of \cite{cel08}, which was based on the historical data. To further understand the physical processes involved in the active phase of PKS\,2052$-$47, SED modeling is needed. The results of this will be presented elsewhere.

\section{Outlook}
The multiwavelength campaign of PKS\,2052$-$47 provides an opportunity to investigate the emission mechanisms and physical processes involved in this radio quasar during its flaring state. When all datasets will be collected, we will be able to compare the VLBI parsec-scale images at 8\,GHz and 22\,GHz with the results from the other wavelengths, especially that of high energies, to probe the source of high-energy emission. Also, the varying broadband SED modeling will be necessary to understand the source in its active phase.

\bigskip % extra skip inserted
\begin{acknowledgments}
We thank to M. B\"ock, C. Ricci, L. Barrag\'an, and J. Wilms for valuable discussions, and D. Thompson for advice and encouragement. This research was supported by the EU Framework 6 Marie Curie Early Stage Training program under contract number MEST/CT/2005/19669 “ESTRELA”. C. S. Chang is a member of the International Max Planck Research School for Astronomy and Astrophysics. This publication is based on data acquired with the Atacama Pathfinder Experiment (APEX), a collaboration between the Max-Planck-Institut f\"ur Radioastronomie, the European Southern Observatory, and the Onsala Space Observatory. This research has been partially funded by the Bundesministerium f\"ur Wirtschaft und Technologie under Deutsches Zentrum f\"ur Luft- und Raumfahrt grant number 50OR0808. This research has made use of the NASA/IPAC Extragalactic Database (NED) which is operated by the Jet Propulsion Laboratory, California Institute of Technology, under contract with the National Aeronautics and Space Administration.%\textbf{Acknowledge Swift!}
\end{acknowledgments}

\bigskip % extra skip inserted

%\bibliographystyle{aa}
%\bibliography{pks2052-47}

\end{document}